\parindent=1cm \parskip=\baselineskip
\raggedbottom
\def\half{{\textstyle{1\over2}}}
\def\pmb#1{\setbox0=\hbox{#1}%
        \kern-.025em\copy0\kern-\wd0
        \kern.05em\copy0\kern-\wd0
        \kern-.025em\raise.0433em\box0 }
\centerline{\bf Coupled Classical and Quantum Oscillators}\par
\vskip 14pt
\centerline{Rachael M.~McDermott\footnote{*}{Current address:
Department of Nuclear Engineering, Massachusetts Institute of Technology,
Cambridge, Massachusetts~~02139}\footnote{${}^1$}{E-mail:  rachmcd@mit.edu}
and Ian H.~Redmount\footnote{${}^2$}{E-mail:  redmouih@slu.edu}}
\vskip 14pt
\centerline{\sl Department of Physics}
\centerline{\sl Parks College of Engineering, Aviation, and Technology}
\centerline{\sl Saint Louis University}
\centerline{\sl 3450 Lindell Boulevard}
\centerline{\sl St.~Louis, Missouri~~63103--1110}
\vskip 14pt
\centerline{(Submitted to {\it Progress of Theoretical Physics\/})}
\vskip 14pt
\centerline{\bf ABSTRACT}\par\nobreak
Some of the most enduring questions in physics---including the quantum
measurement problem and the quantization of gravity---involve the
interaction of a quantum system with a classical environment.  Two
linearly coupled harmonic oscillators provide a simple, exactly soluble
model for exploring such interaction.  Even the ground state of a pair
of identical oscillators exhibits effects on the quantum nature of one
oscillator, e.g., a diminution of position uncertainty, and an increase
in momentum uncertainty and uncertainty product, from their unperturbed
values.  Interaction between quantum and classical oscillators is
simulated by constructing a quantum state with one oscillator initially
in its ground state, the other in a coherent or Glauber state.  The
subsequent wave function for this state is calculated exactly, both for
identical and distinct oscillators.  The reduced probability distribution
for the quantum oscillator, and its position and momentum expectation
values and uncertainties, are obtained from this wave function.  The
oscillator acquires an oscillation amplitude corresponding to a beating
between the normal modes of the system; the behavior of the position and
momentum uncertainties can become quite complicated.  For oscillators with
equal unperturbed frequencies, i.e., at resonance, the uncertainties exhibit
a time-dependent quantum squeezing which can be extreme.
\vskip 14pt
\centerline{PACS number(s):  03.65.-w, 04.60.-m}
\vskip\baselineskip
\centerline{\bf I.~~INTRODUCTION}\par\nobreak
Interaction of the quantum and classical worlds---i.e., between a system
for which quantum effects predominate and a system operating in a classical
limit---lies at the heart of some of the most profound and vexing problems
in physics.  The quantum measurement problem, of course, is one of the
central mysteries of quantum physics; it has engendered concepts from
Schr\"odinger's Cat and the many-worlds interpretation to consistent histories.
The search for a quantum theory of gravitation also entails linking quantum
matter sources to apparently classical spacetime geometry, as the Einstein
field equations display.  While now largely taken for granted, the notion
that there must exist a quantum version of gravitation was at one time
the subject of some debate~[1], and even of experimental test~[2].  One
longstanding argument that gravitation must be a quantum phenomenon with a
classical limit, and not a strictly classical field, is that an ordinary
quantum system coupled to a truly classical field would ``radiate away'' its
quantum nature---commutator values, uncertainties, etc.---leading to observable
violations of quantum mechanics.  A pair of simple harmonic oscillators
with linear coupling provides an ideal test bed for exploring this aspect
of quantum/classical interaction.

Two linearly coupled quantum-mechanical simple harmonic oscillators, e.g.,
two masses on springs, connected by a third spring, constitute an ideally
simple model of quantum/classical coupling.  First, the system separates into
normal modes behaving as independent oscillators, so the evolution of the
system from any initial data can be followed exactly.  Second, the classical
limit of a quantum oscillator is easily described by a coherent or Glauber
state~[3,4]:  a Gaussian wave packet of fixed width, the centroid of which
follows a classical trajectory.  In the present work we analyze the behavior
of coupled oscillators with one initially in its quantum ground state, the
other initially in such a coherent state.  The problem of a quantum oscillator
coupled to an actual classical, i.e., $c$-number oscillator calls for
different techniques and will be examined in a subsequent paper.

The simplest version of the two-oscillator system consists of two identical
oscillators, with equal masses, spring constants, and frequencies, plus a
connecting spring with its own spring constant.  Even in the simplest of
quantum states---its ground state---this system reveals interesting effects
on the oscillators' dynamics:  The position uncertainty of an individual
oscillator is reduced below its uncoupled value, while its momentum uncertainty
is increased.  (Hence, the second oscillator does not act as a ``heat bath,''
which would increase both uncertainties.)  The product of the uncertainties
is increased; the Uncertainty Principle is not violated.  If the system is
started with one oscillator in its quantum ground state and the other in
a quasi-classical coherent state, e.g., if the coupling between them is
``turned on'' at some initial time, then subsequently the two normal modes
evolve one in a coherent state, the other in a modified state termed a
displaced squeezed state, e.g., in quantum optics and the analysis
of gravitational-wave detectors~[4,5].  The initially quantum
oscillator acquires an oscillating position expectation value---a ``beat''
between the normal modes.  Its position uncertainty oscillates through
values below its uncoupled, ground-state value, while its momentum
uncertainty oscillates above its uncoupled value.  The product of these
uncertainties oscillates through values above~$\hbar/2$, never violating
the Uncertainty Principle.

The general two-oscillator system, with arbitrary masses, spring constants,
and frequencies, can likewise be analyzed exactly.  We construct the wave
function for a state with one oscillator initially in its ground state, the
other in a coherent state.  From this we obtain position and momentum
expectation values and uncertainties for the first oscillator.  Expectation
values behave very like those in the identical-oscillator (``symmetric'')
case, but the expressions we find for uncertainties in the fully general
case are rather opaque.  However, one simple case may be of particular
physical importance:  that of oscillators with different masses, but
equal (uncoupled) frequencies, i.e., that of quantum and classical
oscillators interacting ``at resonance.''  In that case quantum uncertainties
clearly exhibit the behaviors found in the symmetric case:  Position
uncertainty is reduced from its uncoupled quantum value--toward the
classical zero value--while momentum uncertainty is increased.  The
product of these remains always above the Uncertainty Principle bound;
the system of course never actually violates quantum mechanics.

The simplest case of two identical oscillators is detailed in Sec.~II
below.  The fully general case is shown in Sec.~III; the special but
physically important equal-frequency case is treated in Sec.~IV.  We
summarize our results and conclusions in Sec.~V.  For ease of comparison
between quantum and classical regimes, we retain the constant~$\hbar$
explicitly throughout.
\vskip\baselineskip
\centerline{\bf II.~~SYMMETRIC COUPLED OSCILLATORS}\par\nobreak
The simplest system of two coupled harmonic oscillators consists of two
identical, one-dimensional oscillators with linear coupling.  For example,
two identical masses~$m$, each on a linear spring with spring constant~$k$
[so each oscillator has uncoupled angular frequency $\omega=(k/m)^{1/2}$],
connected by a third spring with spring constant~$\kappa$, constitute such
a system.
\vskip\baselineskip
\centerline{\bf A.~~Hamiltonian and normal modes.}\par\nobreak
The Hamiltonian for this system can be written
$$\openup 2\jot
\eqalignno{H&={p_1^2+p_2^2\over2m}+\half kx_1^2+\half kx_2^2+
\half\kappa(x_2-x_1)^2&(2.1{\rm a})\cr
&={p_+^2+p_-^2\over2m}+\half kx_+^2+\half(k+2\kappa)x_-^2
\ ,&(2.1{\rm b})\cr}$$
in terms of individual oscillator displacements (from equilibrium)~$x_1$
and~$x_2$ and momenta~$p_1$ and~$p_2$, or normal-mode coordinates~$x_\pm$
and momenta~$p_\pm$.  The normal-mode amplitudes are related to the
individual coordinates via
$$\openup 2\jot
\eqalignno{x_+&={x_1+x_2\over\sqrt{2}}&(2.2{\rm a})\cr
x_-&={x_2-x_1\over\sqrt{2}}\ ,&(2.2{\rm b})\cr}$$
where the scale factor is chosen to make this a unitary transformation.
The normal-mode momenta~$p_\pm$, of course, are conjugate to these.  The
separated form~(2.1b) of the Hamiltonian allows the system to be treated
as two independent harmonic oscillators.
\vskip\baselineskip
\centerline{\bf B.~~Ground state.}\par\nobreak
The quantum ground-state wave function of the system can be written as the
product of the normal-mode ground states, viz.,
$$\Psi_0(x_\pm,t)=\left({m^2\omega\Omega\over\pi^2\hbar^2}\right)^{1/4}
\exp\left(-{m\omega\over2\hbar}\,x_+^2\right)\,
\exp\left(-{m\Omega\over2\hbar}\,x_-^2\right)\,
\exp\left[-{\textstyle{i\over2}}(\omega+\Omega)t\right]\ .\eqno(2.3)$$
The normal-mode angular frequencies follow from Hamiltonian~(2.1b):
$$\openup 2\jot
\eqalignno{\omega&=\left({k\over m}\right)^{1/2}&(2.4{\rm a})\cr
\noalign{\hbox{for the $+$~mode, and}}
\Omega&=\left({k+2\kappa\over m}\right)^{1/2}={\omega\over\gamma}\ ,
&(2.4{\rm b})\cr}$$
for the $-$~mode, where the last expression defines the parameter~$\gamma$.

The probability distribution for the position of one oscillator, with the
system in this state, is obtained by integrating the probability density
over the other oscillator coordinate.  The result is
$$\openup 2\jot
\eqalignno{{\cal P}_1(x_1,t)&=\int_{-\infty}^\infty\Psi_0^*(x_\pm,t)\,
\Psi_0(x_\pm,t)\,dx_2&\cr
&=\left({m^2\omega\Omega\over\pi^2\hbar^2}\right)^{1/2}\int_{-\infty}^\infty
\exp\left(-{m\omega\over2\hbar}(x_1+x_2)^2-{m\Omega\over2\hbar}(x_2-x_1)^2
\right)\,dx_2&\cr
&=\left({2m\omega\over\pi\hbar(1+\gamma)}\right)^{1/2}
\,\exp\left(-{2m\omega\over\hbar(1+\gamma)}\,x_1^2\right)\ .&
(2.5)\cr}$$
This is a Gaussian probability distribution with standard deviation
$$\sigma_x^{(1)}=\sqrt{\left({\hbar\over2m\omega}\right)\,{1+\gamma\over2}}\ .
\eqno(2.6)$$
The unperturbed vacuum-state probability distribution for the position~$x$
of a single oscillator is
$${\cal P}_0(x,t)=\left({m\omega\over\pi\hbar}\right)^{1/2}\,\exp\left(
-{m\omega\over\hbar}\,x^2\right)\ ,\eqno(2.7{\rm a})$$
with standard deviation
$$\sigma_x^{(0)}=\sqrt{\hbar\over2m\omega}\ .\eqno(2.7{\rm b})$$
With $\Omega>\omega$, i.e., $\gamma<1$, the effect of the coupling is to
{\it reduce\/} the position uncertainty of the oscillator.

The probability distribution for the momentum of one oscillator can be
calculated similarly, or by utilizing the symmetry of a harmonic oscillator
under the interchange $\sqrt{m\omega}\,x\leftrightarrow p/\sqrt{m\omega}$.
The result is
$${\cal P}_1(p_1,t)=\left({2\gamma\over\pi\hbar m\omega(1+\gamma)}\right)^{1/2}
\,\exp\left(-{2\gamma\over\hbar m\omega(1+\gamma)}\,p_1^2\right)\ ,
\eqno(2.8{\rm a})$$
with standard deviation
$$\sigma_p^{(1)}=\sqrt{\left({\hbar m\omega\over2}\right)\,{1+\gamma\over
2\gamma}}\ .\eqno(2.8{\rm b})$$
These can be compared with the unperturbed results
$$\openup 2\jot
\eqalignno{{\cal P}_0(p,t)&=\left({1\over\pi\hbar m\omega}\right)^{1/2}\,
\exp\left(-{1\over\hbar m\omega}\,p^2\right)&(2.9{\rm a})\cr
\sigma_p^{(0)}&=\sqrt{\hbar m\omega\over2}\ .&(2.9{\rm b})\cr}$$
The coupling {\it increases\/} the momentum uncertainty of the individual
oscillator.

Thus the net effects of the coupling, in the ground state of the coupled
system, are:
\item{$\bullet$}To decrease the position uncertainty of the individual 
oscillator;
\item{$\bullet$}To increase the momentum uncertainty of the oscillator;
\item{$\bullet$}To increase its uncertainty product $\sigma_x\sigma_p$
above the minimum value~$\hbar/2$;
\item{$\bullet$}To break the symmetry between $\sqrt{m\omega}\,x$ and
$p/\sqrt{m\omega}$ for the individual oscillator.\par
\noindent The last is not so unexpected, since the coupling perturbation is
in position and not symmetrically in momentum.

The effects of the coupling {\it cannot\/} be characterized by the introduction
of a finite temperature for the individual oscillator.  That is, the second
oscillator does not act as a ``heat bath,'' even though coupling to a large
number of oscillators should do so.  The probability distributions and
uncertainties for an oscillator in thermal equilibrium at temperature~$T$
are~[6]:
$$\openup 2\jot
\eqalignno{{\cal P}_T(x)&=\left({m\omega\over\pi\hbar}\,\tanh(\beta/2)
\right)^{1/2}\,\exp\left(-{m\omega\over\hbar}\,\tanh(\beta/2)\,x^2\right)
&(2.10{\rm a})\cr
\sigma_x^{(T)}&=\sqrt{{\hbar\over2m\omega}\,\coth(\beta/2)}&(2.10{\rm b})\cr
{\cal P}_T(p)&=\left({\tanh(\beta/2)\over\pi\hbar m\omega}\right)^{1/2}\,
\exp\left(-{\tanh(\beta/2)\over\hbar m\omega}\,p^2\right)&(2.10{\rm c})\cr
\sigma_p^{(T)}&=\sqrt{{\hbar m\omega\over2}\,\coth(\beta/2)}\ ,
&(2.10{\rm d})\cr}$$
with $\beta\equiv\hbar\omega/(k_BT)$.  The net effects of a finite temperature
are to increase both~$\sigma_x$ and~$\sigma_p$, as well as the uncertainty
product~$\sigma_x\sigma_p$, while preserving the symmetry
between~$\sqrt{m\omega}\,x$ and~$p/\sqrt{m\omega}$.
\vskip\baselineskip
\centerline{\bf C.~~Coupled ground/coherent state.}\par\nobreak
With the system in its ground state, both oscillators display fully
quantum dynamics.  We model a quantum oscillator coupled to a classical
oscillator by a different choice of state:  We envision oscillator~\#1 in
its unperturbed ground state, and oscillator~\#2 in a coherent or Glauber
state with classical amplitude~$X_0$, at time $t=0$.  We follow the
subsequent evolution of this state, and determine the probability
distributions, expectation values, and uncertainties for oscillator~\#1.
In effect, oscillator~\#1 is in its ground state, oscillator~\#2 is
behaving classically with oscillation amplitude~$X_0$, and the coupling
``turns on'' at time $t=0$.  For simplicity, we assume the classical
momentum of the second oscillator is zero at $t=0$, i.e., it is at a
maximum of its classical trajectory at that instant.
\vskip\baselineskip
\centerline{\pmb{\it 1.~~Initial wave function.}}\par\nobreak
The initial wave function for the system, i.e., the wave function at
time $t=0$, for this coupled ground/coherent state is~[3,4]:
$$\openup 2\jot
\eqalignno{\Psi(x_1,x_2,0)&=\left({m\omega\over\pi\hbar}\right)^{1/2}
\,\exp\left(-{m\omega\over2\hbar}\,x_1^2\right)\,
\exp\left(-{m\omega\over2\hbar}(x_2-X_0)^2\right)\ .&(2.11{\rm a})\cr
\noalign{\hbox{In terms of the normal-mode amplitudes, this is simply}}
\Psi(x_\pm,0)&=\left({m\omega\over\pi\hbar}\right)^{1/2}
\,\exp\left[-{m\omega\over2\hbar}\left(x_+-{X_0\over\sqrt{2}}\right)^2\right]
\,\exp\left[-{m\omega\over2\hbar}\left(x_--{X_0\over\sqrt{2}}\right)^2\right]
\ .&\cr&&(2.11{\rm b})\cr}$$
This corresponds to a coherent state of the $+$~mode, with initial
amplitude~$X_0/\sqrt{2}$ and momentum zero.  The initial state of the
$-$~mode, however, is a displaced squeezed state~[4,5]:  The initial
displacement is $X_0/\sqrt{2}$, the initial momentum is zero, but the
initial width of the Gaussian wave packet is proportional to~$1/\sqrt{\omega}$
rather than the $1/\sqrt{\Omega}$ required for a coherent state of
this mode.  (In fact this is a displaced ``antisqueezed'' state,
with $1/\sqrt{\omega}>1/\sqrt{\Omega}$.)
\vskip\baselineskip
\centerline{\pmb{\it 2.~Propagation of the wave function}}\par\nobreak
The wave function for all times $t\ge0$ is obtained by applying
harmonic-oscillator propagators~[7] {\it for the normal modes\/} to the
initial wave function~$\Psi(x_\pm,0)$.  The result is the double integral
$$\openup 2\jot
\eqalignno{\Psi(y_\pm,t)=&\left({m\omega\over\pi\hbar}\right)^{1/2}
\left({m\omega\over2\pi i\hbar\sin(\omega t)}\right)^{1/2}
\left({m\Omega\over2\pi i\hbar\sin(\Omega t)}\right)^{1/2}&\cr
&\qquad\times\int_{-\infty}^\infty\int_{-\infty}^\infty
\exp\left[-{m\omega\over2\hbar}\left(x_+-{X_0\over\sqrt{2}}\right)^2\right]\,
\exp\left[-{m\omega\over2\hbar}\left(x_--{X_0\over\sqrt{2}}\right)^2\right]&\cr
&\qquad\qquad\qquad\times\exp\left({im\omega\over2\hbar\sin(\omega t)}
[(y_+^2+x_+^2)\cos(\omega t)-2x_+y_+]\right)&\cr
&\qquad\qquad\qquad\times\exp\left({im\Omega\over2\hbar\sin(\Omega t)}
[(y_-^2+x_-^2)\cos(\Omega t)-2x_-y_-]\right)\,dx_+\,dx_-\ ,&\cr&&(2.12)\cr}$$
where coordinates~$y$, at arbitrary time~$t$, are distinguished from
coordinates~$x$ at $t=0$.  The integral over~$x_+$ gives a coherent state
in~$y_+$, with classical amplitude $(X_0/\sqrt{2})\cos(\omega t)$ and
momentum $-m\omega(X_0/\sqrt{2})\sin(\omega t)$.  The integral over~$x_-$
gives a displaced squeezed state in~$y_-$.  Ultimately, the wave function
takes the form
$$\openup 2\jot
\eqalignno{\Psi(y_\pm,t)&=\psi_c(y_+,t)\,\psi_{ds}(y_-t)\ ,&(2.13{\rm a})\cr
\noalign{\hbox{with~[3,4]}}
\psi_c(y_+,t)&=\left({m\omega\over\pi\hbar}\right)^{1/4}\,
\exp\left(-{i\omega t\over2}\right)\,
\exp\left({im\omega\over4\hbar}X_0^2\sin(\omega t)\,\cos(\omega t)\right)&\cr
&\qquad\qquad\times\exp\left(-{im\omega\over\hbar}{X_0\over\sqrt{2}}
\sin(\omega t)\,y_+\right)\,\exp\left[-{m\omega\over2\hbar}
\left(y_+-{X_0\over\sqrt{2}}\cos(\omega t)\right)^2\right]&\cr
&&(2.13{\rm b})\cr
\noalign{\hbox{and~[4,5]}}
\psi_{ds}(y_-,t)&=\left({m\omega\over\pi\hbar}\right)^{1/4}\left(
{\cos(\Omega t)-i\gamma\sin(\Omega t)\over\cos^2(\Omega t)+
\gamma^2\sin^2(\Omega t)}\right)^{1/2}\,
\exp\left({im\omega\gamma X_0^2\cos(\Omega t)\,\sin(\Omega t)\over
4\hbar[\cos^2(\Omega t)+\gamma^2\sin^2(\Omega t)]}\right)&\cr
&\qquad\times\exp\left({-im\omega\gamma(X_0/\sqrt{2})\sin(\Omega t)\,y_-\over
\hbar[\cos^2(\Omega t)+\gamma^2\sin^2(\Omega t)]}\right)\,
\exp\left({-im\Omega(1-\gamma^2)\cos(\Omega t)\,\sin(\Omega t)\,
y_-^2\over2\hbar[\cos^2(\Omega t)+\gamma^2\sin^2(\Omega t)]}\right)&\cr
&\qquad\times\exp\left[{-m\omega\over2\hbar[\cos^2(\Omega t)+
\gamma^2\sin^2(\Omega t)]}\,\left(y_--{X_0\over\sqrt{2}}\,\cos(\Omega t)
\right)^2\right]\ ,&(2.13{\rm c})\cr}$$
in terms of normal-mode coordinates~$y_\pm$ and time $t\ge0$.
\vskip\baselineskip
\centerline{\pmb{\it 3.~~Reduced probability distribution for~$y_1$.}}
\par\nobreak
As before, the probability distribution for the single oscillator
coordinate~$y_1$ is obtained by integrating the full probability density
over~$y_2$.  We obtain
$$\openup 2\jot
\eqalignno{{\cal P}(y_1,t)&=\int_{-\infty}^\infty|\psi_c(y_+,t)\,
\psi_{ds}(y_-,t)|^2\,dy_2&\cr
&={m\omega\over\pi\hbar}\,\left({1\over\cos^2(\Omega t)
+\gamma^2\sin^2(\Omega t)}\right)^{1/2}&\cr
&\qquad\times\int_{-\infty}^\infty\exp\left[-{m\omega\over\hbar}\,\left(
y_+-{X_0\over\sqrt{2}}\cos(\omega t)\right)^2\right]\,\exp\left(
-{m\omega\over\hbar}{{\displaystyle{\left(y_--{X_0\over\sqrt{2}}\cos(\Omega t)
\right)^2}}\over\cos^2(\Omega t)+\gamma^2\sin^2(\Omega t)}\right)\,dy_2&\cr
&=\left({2m\omega\over\pi\hbar[1+\cos^2(\Omega t)+\gamma^2\sin^2(\Omega t)]}
\right)^{1/2}\,\exp\left(-{2m\omega\over\hbar}\,{\{y_1-\half X_0
[\cos(\omega t)-\cos(\Omega t)]\}^2\over1+\cos^2(\Omega t)+
\gamma^2\sin^2(\Omega t)}\right)&\cr&&\cr
&=\left({2m\omega\over\pi\hbar[1+\cos^2(\Omega t)+\gamma^2\sin^2(\Omega t)]}
\right)^{1/2}\,\exp\left(-{2m\omega\over\hbar}\,{\left[y_1-X_0
\sin\left({\displaystyle{\Omega-\omega\over2}}t\right)\,
\sin\left({\displaystyle{\Omega+\omega\over2}}t\right)
\right]^2\over1+\cos^2(\Omega t)+\gamma^2\sin^2(\Omega t)}\right)\ ,&\cr
&&(2.14)\cr}$$
treating $y_+$ and~$y_-$ as functions of forms~(2.2a) and~(2.2b) of
coordinates~$y_1$ and~$y_2$.
\vskip\baselineskip
\centerline{\pmb{\it 4.~~Quantum expectation values and uncertainties
for oscillator~\#1.}}\par\nobreak
The expectation value and uncertainty for the position of oscillator~\#1
can be read directly from the Gaussian distribution~${\cal P}(y_1,t)$.
We find
$$\openup 2\jot
\eqalignno{\langle y_1\rangle&={X_0\over2}\,[\cos(\omega t)-\cos(\Omega t)]
=X_0\,\sin\left({\Omega-\omega\over2}t\right)\,\sin\left({\Omega+\omega\over2}t
\right)&(2.15{\rm a})\cr
\noalign{\hbox{and}}
\sigma_y^{(1)}&=\sqrt{{\hbar\over4m\omega}\,[1+\cos^2(\Omega t)+
\gamma^2\sin^2(\Omega t)]}\ .&(2.15{\rm b})\cr}$$
The behavior of~$\langle y_1\rangle$ can be quite complicated; it can be
as large as~$X_0$ or as negative as~$-X_0$.  In the case of {\it weak
coupling,\/} i.e., $\kappa\ll k$ or $\Omega-\omega\ll\omega$, it
oscillates with a slowly oscillating amplitude, exhibiting ``beats''
between the normal modes.  The position uncertainty~$\sigma_y^{(1)}$
oscillates between the unperturbed value $\sqrt{\hbar/(2m\omega)}$ and the
{\it smaller\/} value $\sqrt{\hbar(1+\gamma^2)/(4m\omega)}$.  

Momentum expectation values and uncertainties cannot be determined from
the reduced probability distribution~${\cal P}(y_1,t)$; the full wave
function~$\Psi(y_\pm,t)$ is needed.  The normal-mode relations~(2.2a)
and~(2.2b), plus the chain rule, imply
$$\openup 2\jot
\eqalignno{p_1&={p_+-p_-\over\sqrt{2}}\ ,&(2.16{\rm a})\cr
\noalign{\hbox{whence follow}}
\langle p_1\rangle&={\langle p_+\rangle-\langle p_-\rangle\over
\sqrt{2}}&(2.16{\rm b})\cr
\noalign{\hbox{and}}
\sigma_p^{(1)}&=\sqrt{\sigma_p^{(+)2}+\sigma_p^{(-)2}\over2}\ ,
&(2.16{\rm c})\cr}$$
the last since the normal modes are uncorrelated in this state.
The necessary expectation values and uncertainties can be read from
the momentum-space wave functions for the normal modes, viz.,
$$\openup 2\jot
\eqalignno{\tilde{\psi}_c(p_+,t)&=\int_{-\infty}^\infty{e^{-ip_+y_+/\hbar}
\over(2\pi\hbar)^{1/2}}\,\psi_c(y_+,t)\,dy_+&\cr
&=\left({1\over m\omega\pi\hbar}\right)^{1/4}\,\exp\left(-{i\omega t\over2}
\right)\,\exp\left(-{im\omega\over4\hbar}X_0^2\cos(\omega t)\,\sin(\omega t)
\right)&\cr
&\qquad\times\exp\left(-{i\over\hbar}{X_0\over\sqrt{2}}\cos(\omega t)\,p_+
\right)\,\exp\left[-{1\over2m\omega\hbar}\,\left(p_++m\omega{X_0\over\sqrt{2}}
\,\sin(\omega t)\right)^2\right]&\cr&&(2.17{\rm a})\cr
\noalign{\hbox{and}}
\tilde{\psi}_{ds}(p_-,t)&=\int_{-\infty}^\infty{e^{-ip_-y_-/\hbar}
\over(2\pi\hbar)^{1/2}}\,\psi_{ds}(y_-,t)\,dy_-&\cr
&=\left({\gamma\over m\Omega\pi\hbar}\right)^{1/4}\left({\gamma\cos(\Omega t)
-i\sin(\Omega t)\over\gamma^2\cos^2(\Omega t)+\sin^2(\Omega t)}\right)^{1/2}\,
\exp\left({-im\Omega\gamma^2X_0^2\cos(\Omega t)\,\sin(\Omega t)\over
4\hbar[\gamma^2\cos^2(\Omega t)+\sin^2(\Omega t)]}\right)&\cr
&\qquad\times\exp\left({-i\gamma^2(X_0/\sqrt{2})\cos(\Omega t)\,p_-\over
\hbar[\gamma^2\cos^2(\Omega t)+\sin^2(\Omega t)]}\right)\,
\exp\left({i(1-\gamma^2)\cos(\Omega t)\,\sin(\Omega t)\,
p_-^2\over2m\Omega\hbar[\gamma^2\cos^2(\Omega t)+\sin^2(\Omega t)]}\right)&\cr
&\qquad\times\exp\left[{-\gamma\over2m\Omega\hbar[\gamma^2\cos^2(\Omega t)+
\sin^2(\Omega t)]}\,\left(p_-+m\Omega{X_0\over\sqrt{2}}\,\sin(\Omega t)
\right)^2\right]\ .&\cr&&(2.17{\rm b})\cr}$$
These imply
$$\openup 2\jot
\eqalignno{\langle p_+\rangle&=-m\omega{X_0\over\sqrt{2}}\,\sin(\omega t)\ ,
&(2.18{\rm a})\cr
\sigma_p^{(+)}&=\sqrt{m\omega\hbar\over2}\ ,&(2.18{\rm b})\cr
\langle p_-\rangle&=-m\Omega{X_0\over\sqrt{2}}\,\sin(\Omega t)\ ,
&(2.18{\rm c})\cr
\hbox{and}\qquad\sigma_p^{(-)}&=\sqrt{{m\Omega\hbar\over2\gamma}[\gamma^2
\cos^2(\Omega t)+\sin^2(\Omega t)]}\ .&(2.18{\rm d})\cr}$$
Hence, relations~(2.16b) and~(2.16c) yield the results
$$\openup 2\jot
\eqalignno{\langle p_1\rangle&=-{mX_0\over2}[\omega\sin(\omega t)
-\Omega\sin(\Omega t)]&(2.19{\rm a})\cr
\hbox{and}\qquad\sigma_p^{(1)}&=\sqrt{{m\omega\hbar\over4}\,\left(
1+\cos^2(\Omega t)+{1\over\gamma^2}\sin^2(\Omega t)\right)}\ .
&(2.19{\rm b})\cr}$$
The momentum and position expectation values are related ``classically,''
via $\langle p_1\rangle=m\,d\langle y_1\rangle/dt$.  The momentum uncertainty
oscillates between the unperturbed value $\sqrt{m\omega\hbar/2}$ and the
{\it larger\/} value $\sqrt{m\omega\hbar(1+\gamma^{-2})/4}$.

Thus the behavior of oscillator~\#1 is perhaps reminiscent of the suggestion
that a quantum oscillator coupled to a classical one would lose its quantum
nature.  In particular, its position uncertainty is reduced below its
unperturbed ground-state value.  It cannot, however, be reduced arbitrarily
close to the classical zero value.  For any value of the coupling, i.e.,
of~$\gamma$---and independent of the classical amplitude~$X_0$---the
uncertainty~$\sigma_y^{(1)}$ is never less than $1/\sqrt{2}$ of the
unperturbed quantum value.  The effect of the coupling could more aptly
be described as a quantum ``squeezing'' than a loss of quantum nature.
(Of course, since this model remains a quantum-mechanical system no matter
what its state, it could never exhibit an actual violation of quantum
mechanics.)  The uncertainty product for oscillator~\#1 is
$$\openup 2\jot
\eqalignno{\sigma_y^{(1)}\sigma_p^{(1)}&={\hbar\over4}\sqrt{
[1+\cos^2(\Omega t)+\gamma^2\sin^2(\Omega t)]\left(1+\cos^2(\Omega t)
+{1\over\gamma^2}\sin^2(\Omega t)\right)}&\cr
&={\hbar\over2}\sqrt{1+\left({1\over\gamma}-\gamma\right)^2
{1-\cos^4(\Omega t)\over4}}&\cr
&={\hbar\over4}\sqrt{\left({1\over\gamma}+\gamma\right)^2-\left({1\over\gamma}
-\gamma\right)^2\cos^4(\Omega t)}\ .&(2.20)\cr}$$
This oscillates between the values
$${\hbar\over2}\le\sigma_y^{(1)}\sigma_p^{(1)}\le{\hbar\over4}\,
\left({1\over\gamma}+\gamma\right)\ ,\eqno(2.21)$$
always in accord with the Uncertainty Principle.
\vskip\baselineskip
\centerline{\bf III.~~GENERAL COUPLED OSCILLATORS}\par\nobreak
The preceding case of two identical oscillators is rather special.  One might
expect coupled quantum and classical oscillators to have very different masses
and spring constants---including, for example, a quantum oscillator coupled
to a mode of a radiation field.  The above analysis is readily generalized
to the case of oscillators with arbitrary masses~$m_1$ and~$m_2$ and
individual spring constants~$k_1$ and~$k_2$, hence unperturbed angular
frequencies $\omega_1=\sqrt{k_1/m_1}$ and $\omega_2=\sqrt{k_2/m_2}$.
\vskip\baselineskip
\centerline{\bf A.~~Normal modes.}\par\nobreak
The Hamiltonian for this two-oscillator system, coupled as before by a
spring with spring constant~$\kappa$, is
$$H={p_1^2\over2m_1}+{p_2^2\over2m_2}+\half k_1x_1^2+\half k_2x_2^2
+\half\kappa(x_2-x_1)^2\ .\eqno(3.1)$$
To separate this into normal modes, we first rescale the coordinates and
momenta thus:
$$\openup 2\jot
\eqalignno{X_1&=\left({m_1\over m_2}\right)^{1/4}\,x_1\qquad\qquad
X_2=\left({m_2\over m_1}\right)^{1/4}\,x_2&(3.2{\rm a})\cr
P_1&=\left({m_2\over m_1}\right)^{1/4}\,p_1\qquad\qquad
P_2=\left({m_1\over m_2}\right)^{1/4}\,p_2\ .&(3.2{\rm b})\cr}$$
The Hamiltonian becomes
$$\openup 2\jot
\eqalignno{H&={P_1^2\over2\mu}+{P_2^2\over2\mu}+\half\mu\omega_1^2X_1^2
+\half\mu\omega_2^2X_2^2&\cr
&\qquad+\half\kappa\left[\left({m_1\over m_2}\right)^{1/4}X_2-
\left({m_2\over m_1}\right)^{1/4}X_1\right]^2&(3.3{\rm a})\cr
\noalign{\hbox{with geometric-mean mass}}
\mu&\equiv(m_1m_2)^{1/2}\ .&(3.3{\rm b})\cr}$$
Now the rotation of coordinates and momenta
$$\openup 2\jot
\eqalignno{\left(\matrix{x_+\cr x_-\cr}\right)&=\left(
\matrix{\cos\alpha&\sin\alpha\cr-\sin\alpha&\cos\alpha\cr}\right)\,\left(
\matrix{X_1\cr X_2\cr}\right)\ ,&(3.4{\rm a})\cr
\hbox{hence,}\qquad\left(\matrix{p_+\cr p_-\cr}\right)&=\left(
\matrix{\cos\alpha&\sin\alpha\cr-\sin\alpha&\cos\alpha\cr}\right)\,\left(
\matrix{P_1\cr P_2\cr}\right)\ ,&(3.4{\rm b})\cr}$$
transforms the Hamiltonian into
$$H={p_+^2\over2\mu}+\half\mu\omega_+^2x_+^2+{p_-^2\over2\mu}+
\half\mu\omega_-^2x_-^2\ ,\eqno(3.5)$$
given the conditions
$$\openup 2\jot
\eqalignno{\alpha&=\half\arctan\left({2\kappa/\mu\over\omega_2^2-\omega_1^2+
{\displaystyle{{\kappa\over\mu}{m_1-m_2\over\mu}}}}\right)\ ,&(3.6{\rm a})\cr
\omega_+&=\left\{\omega_1^2\cos^2\alpha+\omega_2^2\sin^2\alpha+{\kappa\over\mu}
\left[\left({m_1\over m_2}\right)^{1/4}\sin\alpha-\left({m_2\over m_1}
\right)^{1/4}\cos\alpha\right]^2\right\}^{1/2}\ ,&\cr&&(3.6{\rm b})\cr
\hbox{and}\qquad\omega_-&=\left\{\omega_1^2\sin^2\alpha+\omega_2^2\cos^2\alpha
+{\kappa\over\mu}\left[\left({m_1\over m_2}\right)^{1/4}\cos\alpha
+\left({m_2\over m_1}\right)^{1/4}\sin\alpha\right]^2\right\}^{1/2}\ ,&\cr
&&(3.6{\rm c})\cr}$$
fixing the rotation angle~$\alpha$ and the normal-mode
frequencies~$\omega_\pm$.  With this Hamiltonian, the normal modes~$x_\pm$
evolve as independent harmonic oscillators.  The combined
transformations~(3.2a) and~(3.4a), between~$(x_1,x_2)$ and~$(x_+,x_-)$,
remain a unitary transformation in the quantum-mechanical sense, though
not as a matrix for $m_1\neq m_2$.
\vskip\baselineskip
\centerline{\bf B.~~Coupled ground/coherent state.}\par\nobreak
As above, we model a quantum/classical coupling via a state in which
oscillator~\#1 is in its ground state, and oscillator~\#2 is in a coherent
state, with classical amplitude~$x_0$ and momentum zero, at initial time $t=0$.
The initial wave function is
$$\eqalignno{\Psi(x_1,x_2,0)&=\left({m_1m_2\omega_1\omega_2\over\pi^2\hbar^2}
\right)^{1/4}\,\exp\left(-{m_1\omega_1x_1^2+m_2\omega_2(x_2-x_0)^2\over
2\hbar}\right)\ ,&\cr
&&(3.7{\rm a})\cr}$$
in terms of the original oscillator coordinates, or
$$\eqalignno{\Psi(x_\pm,0)=\left({\mu^2\omega_1\omega_2\over\pi^2\hbar^2}
\right)^{1/4}\exp\Biggl(&-{\mu\over2\hbar}\Bigl[(\omega_1\cos^2\alpha+
\omega_2\sin^2\alpha)x_+^2+(\omega_1\sin^2\alpha+\omega_2\cos^2\alpha)x_-^2&\cr
&\quad+(\omega_2-\omega_1)\sin(2\alpha)x_+x_--2\omega_2X_0(\sin\alpha\,x_++
\cos\alpha\,x_-)+\omega_2X_0^2\Bigr]\Biggr)\ ,&\cr&&(3.7{\rm b})\cr}$$
in terms of normal-mode coordinates, with $X_0\equiv(m_2/m_1)^{1/4}x_0$ a
rescaled classical amplitude.

The time-dependent wave function for this state is again obtained by applying
separate harmonic-oscillator propagators~$G_\pm$ for the two normal modes.
The integrand of the propagation double integral takes the form
$$\openup 2\jot
\eqalignno{G_+G_-\Psi(x_\pm,0)&=\left({\mu^2\omega_1\omega_2\over\pi^2
\hbar^2}\right)^{1/4}\left({\mu^2\omega_+\omega_-\over4\pi^2i^2\hbar^2
\sin(\omega_+t)\sin(\omega_-t)}\right)^{1/2}\exp\left(-{\mu\omega_2X_0^2\over
2\hbar}\right)&\cr
&\qquad\times\exp\left({i\mu\omega_+\cot(\omega_+t)\over2\hbar}\,y_+^2\right)\,
\exp\left({i\mu\omega_-\cot(\omega_-t)\over2\hbar}\,y_-^2\right)&\cr
&\qquad\times\exp\Biggl(-{\mu\over2\hbar}\Bigl\{\Bigl[\omega_1\cos^2\alpha+
\omega_2\sin^2\alpha-i\omega_+\cot(\omega_+t)\Bigr]\,x_+^2&\cr
&\hbox to 1.0in{}+\Bigl[\omega_1\sin^2\alpha+\omega_2\cos^2\alpha-i\omega_
-\cot(\omega_-t)\Bigr]\,x_-^2+(\omega_2-\omega_1)\sin(2\alpha)\,x_+x_-&\cr
&\hbox to 1.0in{}-2\Bigl[\omega_2X_0\sin\alpha-i\omega_+\csc(\omega_+t)\,y_+
\Bigr]\,x_+-2\Bigl[\omega_2X_0\cos\alpha-i\omega_-\csc(\omega_-t)\,y_-\Bigr]
\,x_-\Bigr\}\Biggr)\ .&\cr&&(3.8)\cr}$$
Hence, the propagation integral over~$x_\pm$ is a two-dimensional Gaussian
integral of the form
$$\openup 2\jot
\eqalignno{\int\exp\left(-{\mu\over2\hbar}(A^TSA-D^TA-A^TD)\right)\,d^2x
&=\int\exp\left(-{\mu\over2\hbar}(B^TSB-D^TS^{-1}D)\right)\,d^2\xi&\cr
&=\left({4\pi^2\hbar^2\over\mu^2\det S}\right)^{1/2}\exp\left({\mu\over2\hbar}
\,D^TS^{-1}D\right)\ ,&\cr&&(3.9{\rm a})\cr}$$
with
$$\openup 2\jot
\eqalignno{A&=\left(\matrix{x_+\cr x_-\cr}\right)\ ,&(3.9{\rm b})\cr
B&=A-S^{-1}D=\left(\matrix{\xi_+\cr\xi_-\cr}\right)\ ,&(3.9{\rm c})\cr
D&=\left(\matrix{\omega_2X_0\sin\alpha-i\omega_+\csc(\omega_+t)\,y_+\cr{}\cr
\omega_2X_0\cos\alpha-i\omega_-\csc(\omega_-t)\,y_-\cr}\right)\ ,
&(3.9{\rm d})\cr
\noalign{\hbox{and}}
S&=\left(\matrix{\omega_1\cos^2\alpha+\omega_2\sin^2\alpha-i\omega_+
\cot(\omega_+t)&\half(\omega_2-\omega_1)\sin(2\alpha)\cr{}\cr
\half(\omega_2-\omega_1)\sin(2\alpha)&\omega_1\sin^2\alpha+\omega_2
\cos^2\alpha-i\omega_-\cot(\omega_-t)\cr}\right)\ .&\cr&&(3.9{\rm e})\cr}$$
The wave function is then
$$\openup 2\jot
\eqalignno{\Psi(y_\pm,t)&=\left({\mu^2\omega_1\omega_2\over\pi^2\hbar^2}
\right)^{1/4}\left({\omega_+\omega_-\over i^2\sin(\omega_+t)\,\sin(\omega_-t)
\,\det S}\right)^{1/2}\exp\left(-{\mu\omega_2X_0^2\over2\hbar}\right)&\cr
&\qquad\qquad\times\exp\left({\mu\over2\hbar}(D^TS^{-1}D+iY^T\Omega_2Y)\right)
\ ,&(3.10{\rm a})\cr}$$
with
$$\openup 2\jot
\eqalignno{Y&\equiv\left(\matrix{y_+\cr y_-\cr}\right)&(3.10{\rm b})\cr
\noalign{\hbox{and}}
\Omega_2&\equiv\left(\matrix{\omega_+\cot(\omega_+t)&0\cr
0&\omega_-\cot(\omega_-t)\cr}\right)\ ,&(3.10{\rm c})\cr}$$
for all $t\ge0$.  Rewriting column matrix~(3.9d) as
$$D=-i\Omega_1(Y+iZ)\ ,\eqno(3.11{\rm a})$$
with
$$\openup 2\jot
\eqalignno{\Omega_1&\equiv\left(\matrix{\omega_+\csc(\omega_+t)&0\cr
0&\omega_-\csc(\omega_-t)\cr}\right)&(3.11{\rm b})\cr
\noalign{\hbox{and}}
Z&\equiv\left(\matrix{{\displaystyle{\omega_2\over\omega_+}}X_0\sin\alpha\,
\sin(\omega_+t)\cr
{}\cr{\displaystyle{\omega_2\over\omega_-}}X_0\cos\alpha\,\sin(\omega_-t)\cr}
\right)\ ,&(3.11{\rm c})\cr}$$
puts the wave function into the form
$$\openup 2\jot
\eqalignno{\Psi(Y,t)&=\left({\mu^2\omega_1\omega_2\over\pi^2\hbar^2}
\right)^{1/4}\left({\omega_+\omega_-\over i^2\sin(\omega_+t)\,\sin(\omega_-t)
\,\det S}\right)^{1/2}\exp\left(-{\mu\omega_2X_0^2\over2\hbar}\right)&\cr
&\qquad\qquad\times\exp\left(-{\mu\over2\hbar}\Bigl[(Y^T+iZ^T)\Omega_1S^{-1}
\Omega_1(Y+iZ)-iY^T\Omega_2Y\Bigr]\right)\ ,&\cr&&(3.12)\cr}$$
in terms of the matrices defined above.

To determine probability distributions, expectation values, and uncertainties,
it is useful to separate the argument of the exponential in~$\Psi$ into real
and imaginary terms.  The first matrix of coefficients can be written
$$\Omega_1S^{-1}\Omega_1=U+iV\ ,\eqno(3.13)$$
where $U$ and $V$ are real, symmetric $2\times2$~matrices, with~$U$
nonsingular and positive definite.  The wave function is then
$$\openup 2\jot
\eqalignno{\Psi(Y,t)&=\left({\mu^2\omega_1\omega_2\over\pi^2\hbar^2}
\right)^{1/4}\left({\omega_+\omega_-\over i^2\sin(\omega_+t)\,\sin(\omega_-t)
\,\det S}\right)^{1/2}\exp\left(-{\mu\omega_2X_0^2\over2\hbar}\right)&\cr
&\qquad\times\exp\Biggl(-{\mu\over2\hbar}\Bigl[Y^TUY-Z^TVY-Y^TVZ-Z^TUZ&\cr
&\hbox to 1.0in{}+iY^T(V-\Omega_2)Y+i(Z^TUY+Y^TUZ)-iZ^TVZ\Bigr]\Biggr)&\cr
&=\left({\mu^2\omega_1\omega_2\over\pi^2\hbar^2}\right)^{1/4}
\left({\omega_+\omega_-\over i^2\sin(\omega_+t)\,\sin(\omega_-t)
\,\det S}\right)^{1/2}\exp\left(-{\mu\omega_2X_0^2\over2\hbar}\right)&\cr
&\qquad\times\exp\Biggl(-{\mu\over2\hbar}\Bigl[(Y^T-Z^TVU^{-1})U(Y-U^{-1}VZ)
-Z^TVU^{-1}VZ-Z^TUZ&\cr
&\hbox to 2.0in{}+iY^T(V-\Omega_2)Y+i(Z^TUY+Y^TUZ)-iZ^TVZ\Bigr]\Biggr)\ .&\cr
&&(3.14)\cr}$$
This can be simplified:  Since the matrix~$U$ contains no dependence on the
classical amplitude~$X_0$, normalization of the wave function implies
$$\openup 2\jot
\eqalignno{\omega_2X_0^2&=Z^T(VU^{-1}V+U)Z\ ,&(3.15{\rm a})\cr
\noalign{\hbox{and also}}
\det U&={\omega_1\omega_2\omega_+^2\omega_-^2
\over\sin^2(\omega_+t)\,\sin^2(\omega_-t)|\det S|^2}\ .&(3.15{\rm b})\cr}$$
These can also be verified using the explicit expressions shown in Sec.~D
below.  The first result reduces the wave function to the form
$$\openup 2\jot
\eqalignno{\Psi(Y,t)&=\left({\mu^2\omega_1\omega_2\over\pi^2\hbar^2}
\right)^{1/4}\left({\omega_+\omega_-\over i^2\sin(\omega_+t)\,\sin(\omega_-t)
\,\det S}\right)^{1/2}&\cr
&\qquad\qquad\times\exp\left({i\mu\over2\hbar}\Bigl[Z^TVZ-(Z^TUY+Y^TUZ)
-Y^T(V-\Omega_2)Y\Bigr]\right)&\cr
&\qquad\qquad\times\exp\left(-{\mu\over2\hbar}(Y^T-Z^TVU^{-1})U(Y-U^{-1}VZ)
\right)\ .&(3.16)\cr}$$
This wave function is finite or regular for all~$y_\pm$ and $t\ge0$; the
components of~$U$, $U^{-1}$, $V-\Omega_2$, and~$VZ$ contain no vanishing
denominators or divergent circular functions.  This too can be verified
explicitly using the expressions in Sec.~D below.

The probability distribution for the normal-mode coordinates is
$$\openup 2\jot
\eqalignno{{\cal P}(Y,t)&=|\Psi(Y,t)|^2&\cr
&=\left({\mu^2\det U\over\pi^2\hbar^2}\right)^{1/2}
\exp\left(-{\mu\over\hbar}(Y-U^{-1}VZ)^TU(Y-U^{-1}VZ)\right)\ ,&\cr
&&(3.17)\cr}$$
a Gaussian distribution in two dimensions.  To obtain the reduced probability
distribution for the single oscillator coordinate~$y_1$, the inverse of the
transformation to normal-mode coordinates,
$$\openup 2\jot
\eqalignno{Y&=\left(\matrix{\cos\alpha&\sin\alpha\cr-\sin\alpha&\cos\alpha}
\right)\left(\matrix{(m_1/m_2)^{1/4}&0\cr0&(m_2/m_1)^{1/4}\cr}\right)
\left(\matrix{y_1\cr y_2\cr}\right)&\cr
&\equiv N{\cal Y}\ ,&(3.18)\cr}$$
is needed.  Here ${\cal Y}$ denotes the column containing the oscillator
coordinates, and $N$ the product of the two square matrices.  The
matrix~$N$ has unit determinant, but is not orthogonal if masses~$m_1$
and~$m_2$ are unequal.  Distribution~(3.17) can be expressed thus:
$$\openup 2\jot
\eqalignno{{\cal P}({\cal Y},t)&=\left({\mu^2\det U\over\pi^2\hbar^2}
\right)^{1/2}\exp\left(-{\mu\over\hbar}({\cal Y}-N^{-1}U^{-1}VZ)^TN^TUN
({\cal Y}-N^{-1}U^{-1}VZ)\right)&\cr
&&(3.19)\cr}$$
as a distribution for~$y_1$ and~$y_2$.  Reduction to~${\cal P}(y_1,t)$
requires an integral of the general form
$$\openup 2\jot
\eqalignno{\int_{-\infty}^\infty\exp\left[-(\matrix{x_1&x_2\cr})\left(
\matrix{a&c\cr c&b\cr}\right)\left(\matrix{x_1\cr x_2\cr}\right)\right]\,dx_2
&=\int_{-\infty}^\infty\exp(-ax_1^2-2cx_1x_2-bx_2^2)\,dx_2&\cr
&=\left({\pi\over b}\right)^{1/2}\exp\left(-{ab-c^2\over b}\,x_1^2\right)
\ .&(3.20)\cr}$$
Hence, integrating ${\cal P}({\cal Y},t)$ over~$y_2$ yields
$$\openup 2\jot
\eqalignno{{\cal P}(y_1,t)&=\left({\mu\det U\over\pi\hbar(N^TUN)_{22}}
\right)^{1/2}\exp\left(-{\mu\det U\over\hbar(N^TUN)_{22}}\Bigl[
y_1-(N^{-1}U^{-1}VZ)_1\Bigr]^2\right)\ ,&\cr&&(3.21)\cr}$$
a normalized, Gaussian distribution for~$y_1$.
\vskip\baselineskip
\centerline{\bf C.~~Expectation values and uncertainties.}\par\nobreak
The expectation value and uncertainty for the position of oscillator~\#1 in
this state can be read off of~${\cal P}(y_1,t)$.  They are
$$\openup 2\jot
\eqalignno{\langle y_1\rangle&=(N^{-1}U^{-1}VZ)_1&(3.22{\rm a})\cr
\hbox{and}\qquad\sigma_y^{(1)}&=\sqrt{{\hbar\over2\mu}\,{(N^TUN)_{22}\over
\det U}}\ ,&(3.22{\rm b})\cr}$$
in terms of the above-defined matrices.

To calculate the momentum expectation value and uncertainty, the full wave
function~(3.16) must be used.  First, we obtain
$$\openup 2\jot
\eqalignno{\langle p_1\rangle&=\int\Psi^*{\hbar\over i}\,{\partial\over
\partial y_1}\Psi\,dy_1\,dy_2&\cr
\noalign{\vfil\eject}
&={i\mu\over2}\int\Biggl\{(\matrix{1&0\cr})N^TUN({\cal Y}-N^{-1}U^{-1}VZ)
+({\cal Y}-N^{-1}U^{-1}VZ)^TN^TUN\left(\matrix{1\cr0\cr}\right)&\cr
&\qquad\qquad+i\left[Z^TUN\left(\matrix{1\cr0\cr}\right)
+(\matrix{1&0\cr})N^TUZ\right]&\cr
&\qquad\qquad+i\left[(\matrix{1&0\cr})N^T(V-\Omega_2)N{\cal Y}
+{\cal Y}^TN^T(V-\Omega_2)N\left(\matrix{1\cr0\cr}\right)\right]\Biggr\}
\,{\cal P}({\cal Y},t)\,dy_1\,dy_2&\cr
&=-\mu\left\{N^T\left[U+(V-\Omega_2)U^{-1}V\right]Z\right\}_1\ ,&(3.23)\cr}$$
this last following from the matrix expectation value
$\langle{\cal Y}\rangle=N^{-1}U^{-1}VZ$.  Second, we find
$$\openup 2\jot
\eqalignno{\langle p_1^2\rangle&=-\hbar^2\int\Psi^*\,{\partial^2\over
\partial y_1^2}\,\Psi\,dy_1\,dy_2&\cr
&=-\hbar^2\Biggl[-{\mu\over\hbar}\left\langle(\matrix{1&0\cr})N^TUN\left(
\matrix{1\cr0\cr}\right)+i(\matrix{1&0\cr})N^T(V-\Omega_2)N\left(
\matrix{1\cr0\cr}\right)\right\rangle&\cr
&\qquad\qquad+{\mu^2\over\hbar^2}\left\langle\left\{\left[N^TUN
({\cal Y}-N^{-1}U^{-1}VZ)\right]_1+i\left[N^T(V-\Omega_2)N{\cal Y}\right]_1
+i(N^TUZ)_1\right\}^2\right\rangle\Biggr]&\cr
&=\mu\hbar\left\{N^T[U+i(V-\Omega_2)]N\right\}_{11}&\cr
&\quad-\mu^2\left\langle({\cal Y}-N^{-1}U^{-1}VZ)^TN^T[U+i(V-\Omega_2)]N
\left(\matrix{1&0\cr0&0\cr}\right)N^T[U+i(V-\Omega_2)]N
({\cal Y}-N^{-1}U^{-1}VZ)\right\rangle&\cr
&\qquad\qquad+\mu^2\left(\left\{N^T\left[U+(V-\Omega_2)U^{-1}V\right]Z
\right\}_1\right)^2\ ,&(3.24)\cr}$$
using the facts the square of the first component of any
column matrix~$C$ can always be written
$${C_1}^2=C^T\left(\matrix{1&0\cr0&0\cr}\right)C\ ,\eqno(3.25)$$
and that terms linear in ${\cal Y}-N^{-1}U^{-1}VZ$ appearing in the original
squared expression have zero expectation value.  The last term in
Eq.~(3.24) is precisely $\langle p_1\rangle^2$---without specifying the
matrices---leaving
$$\openup 2\jot
\eqalignno{\sigma_p^{(1)2}&=\langle p_1^2\rangle-\langle p_1\rangle^2&\cr
&=\mu\hbar\left\{N^T[U+i(V-\Omega_2)]N\right\}_{11}&\cr
&\qquad-\mu^2\left\langle({\cal Y}-\langle{\cal Y}\rangle)^T
N^T[U+i(V-\Omega_2)]N\left(\matrix{1&0\cr0&0\cr}\right)N^T[U+i(V-\Omega_2)]N
({\cal Y}-\langle{\cal Y}\rangle)\right\rangle\ .&\cr
&&(3.26)\cr}$$
The last expectation value is the integral of a quadratic form times the
normalized Gaussian~${\cal P}({\cal Y},t)$.  The general form of such an
integral is
$$\left({\det B\over\pi^n}\right)^{1/2}\int X^TAX\,e^{-X^TBX}\,d^nx
=\half\,{\rm Tr}(AB^{-1})\ ,\eqno(3.27)$$
for positive-definite $B$, as can be proved by diagonalizing~$B$ or by
explicit integration of component expressions.  This implies
$$\openup 2\jot
\eqalignno{\sigma_p^{(1)2}&=\mu\hbar\left\{N^T[U+i(V-\Omega_2)]N\right\}_{11}
&\cr
&\qquad-{\mu^2\over2}{\rm Tr}\left[N^T[U+i(V-\Omega_2)]N\left(
\matrix{1&0\cr0&0\cr}\right)N^T[U+i(V-\Omega_2)]N\left({\mu\over\hbar}
N^TUN\right)^{-1}\right]&\cr
&={\mu\hbar\over2}\left\{N^T[U+(V-\Omega_2)U^{-1}(V-\Omega_2)]N\right\}_{11}\ .
&(3.28)\cr}$$
As expected, this is real and of order~$\hbar$, even without using the
specific forms of~$U$, $N$, $V$, or~$\Omega_2$.  It implies the momentum
uncertainty
$$\sigma_p^{(1)}=\sqrt{{\mu\hbar\over2}\left\{N^T[U+(V-\Omega_2)U^{-1}
(V-\Omega_2)]N\right\}_{11}}\ .\eqno(3.29)$$
The probability distribution for momentum~$p_1$ is a Gaussian, with
expectation value~$\langle p_1\rangle$ and standard deviation~$\sigma_p^{(1)}$.
\vskip\baselineskip
\centerline{\bf D.~~Explicit expressions.}\par\nobreak
Explicit expressions for the  wave function, probability distributions,
expectation values, and uncertainties in the coupled ground/coherent state---in
terms of the individual oscillator masses, frequencies, and coupling---involve
an unwieldy number of terms, even with the aid of a computer.  It facilitates
matters to define the following real and imaginary parts, in terms of the
matrices in Eqs.~(3.9e), (3.10c), and~(3.11b): 
$$M=\Omega_1^{-1}S\Omega_1^{-1}={\cal R}-i{\cal I}\ ,\eqno(3.30{\rm a})$$
with real matrices
$$\openup 2\jot
\eqalignno{{\cal R}&=\left(\matrix{\omega_c\rho^2&
\delta\sin\alpha\,\cos\alpha\,\rho\eta\cr{}&{}\cr
\delta\sin\alpha\,\cos\alpha\,\rho\eta&\omega_s\eta^2\cr}\right)
&(3.30{\rm b})\cr
\noalign{\hbox{and}}
{\cal I}&=\Omega_1^{-1}\Omega_2\Omega_1^{-1}
=\left(\matrix{\rho\cos(\omega_+t)&0\cr{}&{}\cr
0&\eta\cos(\omega_-t)\cr}\right)\ ,&(3.30{\rm c})\cr}$$
and frequency combinations
$$\openup 2\jot
\eqalignno{\omega_c&\equiv\omega_1\cos^2\alpha+\omega_2\sin^2\alpha\ ,
&(3.30{\rm d})\cr
\omega_s&\equiv\omega_1\sin^2\alpha+\omega_2\cos^2\alpha\ ,&(3.30{\rm e})\cr
\delta&\equiv\omega_2-\omega_1\ ,&(3.30{\rm f})\cr
\rho&\equiv{\sin(\omega_+t)\over\omega_+}\ ,&(3.30{\rm g})\cr
\hbox{and}\qquad\eta&\equiv{\sin(\omega_-t)\over\omega_-}\ .
&(3.30{\rm h})\cr}$$
The inverse matrix~$M^{-1}$ is the matrix in Eq.~(3.13), implying
$$U+iV={1\over{\cal M}}\,(\tilde{\cal R}-i\tilde{\cal I})\ ,\eqno(3.31)$$
with ${\cal M}\equiv\det M$ and tildes denoting adjoint matrices (transposed
matrices of cofactors, not Hermitian conjugates).  The matrices appearing
in the wave function, {\it et cetera,\/} can then be written:
$$\openup 2\jot
\eqalignno{U&={1\over|{\cal M}|^2}[(\Re{\cal M})\tilde{\cal R}
-(\Im{\cal M})\tilde{\cal I}]\ ,&(3.32{\rm a})\cr
U^{-1}&={1\over|{\cal M}|^2\det U}[(\Re{\cal M}){\cal R}-(\Im{\cal M}){\cal I}]
\ ,&(3.32{\rm b})\cr
V&=-{1\over|{\cal M}|^2}[(\Im{\cal M})\tilde{\cal R}
+(\Re{\cal M})\tilde{\cal I}]\ .&(3.32{\rm c})\cr}$$
The real and imaginary parts of~${\cal M}$ are, explicitly,
$$\openup 2\jot
\eqalignno{\Re{\cal M}&=\rho\eta[\omega_1\omega_2\rho\eta-\cos(\omega_+t)
\cos(\omega_-t)]\ ,&(3.33{\rm a})\cr
\hbox{and}\qquad\Im{\cal M}&=-\rho\eta[\omega_s\eta\cos(\omega_+t)+
\omega_c\rho\cos(\omega_-t)]\ .&(3.33{\rm b})\cr}$$
Identity~(3.15b) is equivalent to
$$|{\cal M}|^2\det U=\omega_1\omega_2\rho^2\eta^2=\det{\cal R}\ ,\eqno(3.34)$$
which can be confirmed by direct evaluation.  Finally,
$$({\cal R}-i{\cal I})(\tilde{\cal R}-i\tilde{\cal I})={\cal M}\,{\bf 1}\ ,
\eqno(3.35{\rm a})$$
where {\bf 1\/} denotes the $2\times2$ identity matrix, implies the identities
$$\Re{\cal M}=\det{\cal R}-\det{\cal I}\ ,\eqno(3.35{\rm b})$$
as may also be obtained directly, and
$${\cal I}\tilde{\cal R}+{\cal R}\tilde{\cal I}=-\Im{\cal M}\,{\bf 1}\ .
\eqno(3.35{\rm c})$$
These simplify the calculations considerably.

Evaluation of the position expectation value~$\langle y_1\rangle$ is
straightforward. Results (3.32b) and~(3.32c) and identities~(3.35b) and~(3.35c)
imply
$$U^{-1}V={1\over|{\cal M}|^2\det U}{\cal I}\tilde{\cal R}\ .\eqno(3.36)$$
Explicit evaluation yields
$$\tilde{\cal R}Z=\omega_1\omega_2X_0\rho\eta\left(\matrix{\eta\sin\alpha\cr
{}\cr\rho\cos\alpha\cr}\right)\ .\eqno(3.37)$$
With identity~(3.34) and form~(3.30c), this gives
$$\openup 2\jot
\eqalignno{\langle Y\rangle&=U^{-1}VZ=\left(\matrix{X_0\sin\alpha\,
\cos(\omega_+t)\cr{}\cr X_0\cos\alpha\,\cos(\omega_-t)\cr}\right)\ ,
&(3.38{\rm a})\cr
\noalign{\hbox{whence follows}}
\langle{\cal Y}\rangle&=N^{-1}U^{-1}VZ&\cr
&=\left(\matrix{(m_2/m_1)^{1/4}X_0\sin\alpha\,\cos\alpha\,[\cos(\omega_+t)
-\cos(\omega_-t)]\cr{}\cr(m_1/m_2)^{1/4}X_0[\sin^2\alpha\,\cos(\omega_+t)
+\cos^2\alpha\,\cos(\omega_-t)]\cr}\right)\ .&(3.38{\rm b})\cr}$$
This gives the desired result:
$$\langle y_1\rangle=(m_2/m_1)^{1/2}x_0\sin\alpha\,\cos\alpha\,[\cos(\omega_+t)
-\cos(\omega_-t)]\ ,\eqno(3.39)$$
in terms of the original classical amplitude~$x_0$ and angle~$\alpha$ and
normal-mode frequencies~$\omega_\pm$ from Eqs.~(3.6a--c).

Evaluation of the position uncertainty~$\sigma_y^{(1)}$ requires the matrix
$$N^TUN={1\over|{\cal M}|^2}[(\Re{\cal M})N^T\tilde{\cal R}N
-(\Im{\cal M})N^T\tilde{\cal I}N]\ .\eqno(3.40)$$
Direct evaluation of the matrix elements
$$\openup 2\jot
\eqalignno{(N^T\tilde{R}N)_{22}&=\left({m_2\over m_1}\right)^{1/2}
(\omega_s\eta^2\sin^2\alpha-2\delta\rho\eta\sin^2\alpha\,\cos^2\alpha
+\omega_c\rho^2\cos^2\alpha)\qquad\qquad&(3.41{\rm a})\cr
\noalign{\hbox{and}}
(N^T\tilde{\cal I}N)_{22}&=\left({m_2\over m_1}\right)^{1/2}
[\eta\sin^2\alpha\,\cos(\omega_-t)+\rho\cos^2\alpha\,\cos(\omega_+t)]
&(3.41{\rm b})\cr}$$
enables us to obtain the final result
$$\openup 2\jot
\eqalignno{\sigma_y^{(1)}&=\Biggl\{{\hbar\over2m_1\omega_1}
\Biggl[\cos^2\alpha\,\left({\omega_s\over\omega_2}
\cos^2(\omega_+t)+{\omega_1\omega_c\over\omega_+^2}\sin^2(\omega_+t)\right)\cr
&\qquad\qquad+\sin^2\alpha\,\left({\omega_c\over\omega_2}\cos^2(\omega_-t)
+{\omega_1\omega_s\over\omega_-^2}\sin^2(\omega_-t)\right)\cr
&\qquad\qquad+2{\delta\over\omega_2}\cos^2\alpha\,\sin^2\alpha\left(
\cos(\omega_+t)\cos(\omega_-t)-{\omega_1\omega_2\over\omega_+\omega_-}
\sin(\omega_+t)\sin(\omega_-t)\right)\Biggr]\Biggr\}^{1/2}\ .&\cr
&&(3.42)\cr}$$
This rather complicated result reduces to form~(2.15b) in the equal-mass,
equal-frequency case.

The momentum expectation value~$\langle p_1\rangle$ is obtained from the
matrix
$$\openup 2\jot
\eqalignno{U+(V-\Omega_2)U^{-1}V&={1\over|{\cal M}|^2}
[(\Re{\cal M})\tilde{\cal R}-(\Im{\cal M})\tilde{\cal I}]&\cr
&\qquad-\left({1\over|{\cal M}|^2}[(\Im{\cal M})\tilde{\cal R}
+(\Re{\cal M})\tilde{\cal I}]+\Omega_2\right){1\over|{\cal M}|^2\det U}
{\cal I}\tilde{\cal R}&\cr
&={1\over|{\cal M}|^2\det U}({\bf 1}-\Omega_2{\cal I})\tilde{\cal R}&\cr
&={1\over\omega_1\omega_2\rho^2\eta^2}\left(\matrix{\sin^2(\omega_+t)&0\cr
{}&{}\cr0&\sin^2(\omega_-t)\cr}\right)\tilde{\cal R}\ .&(3.43)\cr}$$
Result~(3.37) then yields the column matrix
$$\openup 2\jot
[U+(V-\Omega_2)U^{-1}V]Z=X_0\left(\matrix{\omega_+\sin(\omega_+t)
\,\sin\alpha\cr{}\cr\omega_-\sin(\omega_-t)\,\cos\alpha\cr}\right)\ ,
\eqno(3.44)$$
hence, 
$$\openup 2\jot
\eqalignno{N^T[U+(V-\Omega_2)U^{-1}V]Z&=X_0\left(\matrix{(m_1/m_2)^{1/4}
\sin\alpha\,\cos\alpha\,[\omega_+\sin(\omega_+t)-\omega_-\sin(\omega_-t)]\cr
{}\cr(m_2/m_1)^{1/4}[\sin^2\alpha\,\omega_+\sin(\omega_+t)+\cos^2\alpha\,
\omega_-\sin(\omega_-t)]\cr}\right)\ .&\cr&&(3.45)\cr}$$
Result~(3.23) takes the form
$$\langle p_1\rangle=-m_1x_0\left({m_2\over m_1}\right)^{1/2}
\sin\alpha\,\cos\alpha\,[\omega_+\sin(\omega_+t)-\omega_-\sin(\omega_-t)]\ .
\eqno(3.46)$$
As in the symmetric case, the expectation values~$\langle y_1\rangle$
and~$\langle p_1\rangle$ are related by $\langle p_1\rangle=m_1\,
d\langle y_1\rangle/dt$.

The momentum uncertainty~$\sigma_p^{(1)}$ is given by Eq.~(3.29).  After some
manipulation, the requisite matrix takes the form
$$\openup 2\jot
\eqalignno{U+(V-\Omega_2)U^{-1}(V-\Omega_2)&={1\over\det{\cal R}}[
({\bf 1}-\Omega_2{\cal I})\tilde{\cal R}-\tilde{\cal R}{\cal I}\Omega_2&\cr
&\qquad\qquad+(\Re{\cal M})\Omega_2{\cal R}\Omega_2
-(\Im{\cal M})\Omega_2{\cal I}\Omega_2]\ .\qquad&(3.47)\cr}$$
Hence, the uncertainty is obtained from the ${}_{11}$~component of
$$\displaylines{N^T[U+(V-\Omega_2)U^{-1}(V-\Omega_2)]N=\hfill\cr
\hfill{1\over\det{\cal R}}\Biggl[N^T\left(\matrix{\sin^2(\omega_+t)&0\cr
0&\sin^2(\omega_-t)\cr}\right)\tilde{\cal R}N
-N^T\tilde{\cal R}\left(\matrix{\cos^2(\omega_+t)&0\cr
0&\cos^2(\omega_-t)\cr}\right)N\cr
\hfill+(\Re{\cal M})N^T\Omega_2{\cal R}\Omega_2N
-(\Im{\cal M})N^T\Omega_2{\cal I}\Omega_2N\Biggr]\ .\qquad\qquad
{\textstyle{(3.48)}}\cr}$$
The components of the individual matrix products on the right-hand side
can be obtained by explicit calculation.  The results are quite long and
unwieldy, even after some trigonometric simplifications.  When they are
combined, some cancellations and further trigonometric combinations leave
the final result
$$\openup 2\jot
\eqalignno{\sigma_p^{(1)}&=\Biggl\{{\hbar m_1\omega_1\over2}\Biggl[\cos^2\alpha
\left({\omega_c\over\omega_1}\cos^2(\omega_+t)+{\omega_s\omega_+^2\over
\omega_1^2\omega_2}\sin^2(\omega_+t)\right)&\cr
&\hbox to 1in{}+\sin^2\alpha\left({\omega_s\over\omega_1}\cos^2(\omega_-t)
+{\omega_c\omega_-^2\over\omega_1^2\omega_2}\sin^2(\omega_-t)\right)&\cr
&\hbox to 1in{}-2{\delta\over\omega_1}\cos^2\alpha\,\sin^2\alpha\,\left(
\cos(\omega_+t)\,\cos(\omega_-t)-{\omega_+\omega_-\over\omega_1\omega_2}
\sin(\omega_+t)\,\sin(\omega_-t)\right)\Biggr]\Biggr\}^{1/2}\ .&\cr
&&(3.49)\cr}$$
This rather complex result agrees with Eq.~(2.19b) in the symmetric case.

Similar matrix manipulations can be used to confirm identity~(3.15a)
explicitly.  They can also be used to prove the finiteness of the
matrices~$U$, $VZ$, $V-\Omega_2$, and~$U^{-1}$ appearing in the wave
function~$\Psi$, thus:  The combinations $UZ+iVZ$ and
$U+i(V-\Omega_2)$ can be written as finite complex matrices
divided by the complex number $\Xi(t)\equiv{\cal M}(t)/[\rho(t)\eta(t)]$.
Hence, finiteness of all components of~$UZ$, $VZ$, $U$, and $V-\Omega_2$ is
assured provided $\Xi$ neither vanishes nor approaches arbitrarily close to
zero at any time.  This number is obtained from Eqs.~(3.33a) and~(3.33b):
$$\openup 2\jot
\eqalignno{\Re\Xi(t)&=-\cos(\omega_+t)\,\cos(\omega_-t)+{\omega_1\omega_2\over
\omega_+\omega_-}\sin(\omega_+t)\,\sin(\omega_-t)&(3.50{\rm a})\cr
\Im\Xi(t)&=-{\omega_s\over\omega_-}\cos(\omega_+t)\,\sin(\omega_-t)
-{\omega_c\over\omega_+}\cos(\omega_-t)\,\sin(\omega_+t)\ .&(3.50{\rm b})\cr}$$
Consequently, $\Xi$ satisfies the identity
$$\openup 2\jot
\eqalignno{\omega_c\omega_s(\Re\Xi)^2+\omega_1\omega_2(\Im\Xi)^2&=
\omega_c\omega_s\cos^2(\omega_+t)\,\cos^2(\omega_-t)
+{\omega_c\omega_s\omega_1^2\omega_2^2\over\omega_+^2\omega_-^2}
\sin^2(\omega_+t)\sin^2(\omega_-t)&\cr
&\qquad+{\omega_1\omega_2\omega_s^2\over\omega_-^2}\cos^2(\omega_+t)\,
\sin^2(\omega_-t)+{\omega_1\omega_2\omega_c^2\over\omega_+^2}\cos^2(\omega_-t)
\,\sin^2(\omega_+t)\ .&\cr&&(3.51)\cr}$$
With
$$\openup 2\jot
\eqalignno{\zeta&\equiv\max\{\omega_c\omega_s,\omega_1\omega_2\}
&(3.52{\rm a})\cr
\hbox{and}\qquad\lambda&\equiv\min\left\{\omega_c\omega_s,{\omega_c\omega_s
\omega_1^2\omega_2^2\over\omega_+^2\omega_-^2},{\omega_1\omega_2\omega_s^2
\over\omega_-^2},{\omega_1\omega_2\omega_c^2\over\omega_+^2}\right\}\ ,
&(3.52{\rm b})\cr}$$
identity~(3.51) implies $\zeta\,|\Xi|^2\ge\lambda$, i.e.,
$$|\Xi|^2\ge{\lambda\over\zeta}>0\ .\eqno(3.53)$$
As required, $\Xi$ is bounded away from zero by a time-independent bound.
Likewise, matrix~$U^{-1}$ can be written in terms of finite matrices
{\it times\/}~$\Xi$.  But with
$$\openup 2\jot
\eqalignno{\zeta^\prime&\equiv\min\{\omega_c\omega_s,\omega_1\omega_2\}
=\omega_1\omega_2&(3.54{\rm a})\cr
\hbox{and}\qquad\lambda^\prime&\equiv\max\left\{\omega_c\omega_s,
{\omega_c\omega_s\omega_1^2\omega_2^2\over\omega_+^2\omega_-^2},
{\omega_1\omega_2\omega_s^2\over\omega_-^2},
{\omega_1\omega_2\omega_c^2\over\omega_+^2}\right\}\ ,&(3.54{\rm b})\cr}$$
identity~(3.51) also implies $\zeta^\prime\,|\Xi|^2\le\lambda^\prime$, i.e.,
$$|\Xi|^2\le{\lambda^\prime\over\zeta^\prime}<\infty\ .\eqno(3.55)$$
With $\Xi$ bounded away from infinity, the finiteness of~$U^{-1}$ is
also assured.

The expectation values~$\langle y_1\rangle$ and~$\langle p_1\rangle$ for
quantum oscillator~\#1 show the same beating between the normal modes in
the general case as in the symmetric case.  But the
uncertainties~$\sigma_y^{(1)}$ and~$\sigma_p^{(1)}$, which incorporate
the effects of the coupling on the quantum character of the oscillator,
are too complicated in the general case to describe easily.  It is more
illuminating to focus on another special case---one of particular
physical significance.  The effects of one oscillator on another will
certainly be most pronounced when the two are at or near resonance.
It will therefore be most useful to focus on the case of two distinct
oscillators, with unequal masses $m_1\neq m_2$ and correspondingly different
spring constants, but with equal (unperturbed) frequencies
$\omega_1=\omega_2$. 
\vskip\baselineskip
\centerline{\bf IV.~~UNEQUAL-MASS OSCILLATORS AT RESONANCE.}\par\nobreak
The behavior of coupled oscillators of distinct masses but equal unperturbed
frequencies is easily obtained from the general results of Sec.~III above.
In this case the oscillator frequencies and frequency combinations
reduce to $\omega_1=\omega_2=\omega_c=\omega_s=\omega$ and $\delta=0$.
\vskip\baselineskip
\noindent{\bf A.~Normal modes.}\par\nobreak
The normal-mode rotation angle~$\alpha$ for this case follows from Eq.~(3.6a),
i.e.,
$$\tan(2\alpha)={2\mu\over m_1-m_2}\ ,\eqno(4.1)$$
which implies
$$\openup 2\jot
\eqalign{\cos(2\alpha)&={m_1-m_2\over m_1+m_2}\cr
\sin(2\alpha)&={2\mu\over m_1+m_2}\ ,\cr}\eqno(4.2)$$
hence
$$\openup 2\jot
\eqalign{\cos\alpha&=\sqrt{m_1\over m_1+m_2}\cr
\sin\alpha&=\sqrt{m_2\over m_1+m_2}\ .\cr}\eqno(4.3)$$
These reduce to $\alpha=\pi/4$ in the symmetric case $m_1=m_2$, as in
Sec.~II above [Eqs.~(2.2a) and~(2.2b)].  Even with unequal masses, at
resonance, angle~$\alpha$ is independent of the oscillator coupling~$\kappa$.

With these results, Eqs.~(3.6b) and~(3.6c) give the normal-mode frequencies
$$\openup 2\jot
\eqalign{\omega_+&=\omega\cr
\omega_-\equiv\Omega&=\sqrt{\omega^2+{\kappa(m_1+m_2)\over\mu^2}}\ .\cr}
\eqno(4.4)$$
The ratio $\gamma\equiv\omega/\Omega$ can be written
$$\gamma=\left(1+{\kappa(m_1+m_2)\over\mu^2\omega^2}\right)^{-1/2}\ .
\eqno(4.5)$$
Here $\Omega$ and $\gamma$ do depend on~$\kappa$.  Since the masses must obey
$m_1+m_2\ge2\mu$, with equality only for $m_1=m_2$, for given $\mu$ and
$\omega$ values frequency~$\Omega$ is greater, and ratio~$\gamma$
is smaller, with unequal masses than in the symmetric case.
\vskip\baselineskip
\noindent{\bf B.~Ground/coherent state:  coefficient matrices.}\par\nobreak
The behavior of the quantum state with oscillator~\#1 initially in its
unperturbed ground state, and oscillator~\#2 initially in a coherent
state with classical amplitude~$x_0$, follows immediately from the
results of Secs.~III.B--D above.  With the frequencies appropriate to this
case, Eq.~(3.9e) gives a coefficient matrix~$S$ identical in form to that
for the symmetric case, i.e., corresponding to wave function~(2.13a--c).
Matrices~$\Omega_1$ and~$\Omega_2$ are likewise the same as in the symmetric
case; hence, so are $U$, $U^{-1}$, and~$V$.  The only matrices
different in this case are
$$N={1\over\sqrt{\mu(m_1+m_2)}}\left(\matrix{m_1&m_2\cr-\mu&\mu\cr}\right)
\ ,\eqno(4.6{\rm a})$$
its transpose~$N^T$, its inverse
$$N^{-1}={1\over\sqrt{\mu(m_1+m_2)}}\left(\matrix{\mu&-m_2\cr\mu&m_1\cr}\right)
\ ,\eqno(4.6{\rm b})$$
and the column matrix
$$Z=x_0\,\sqrt{\mu\over m_1+m_2}\,\left(\matrix{\sqrt{m_2/m_1}\,
\sin(\omega t)\cr{}\cr\gamma\,\sin(\Omega t)\cr}\right)\ .\eqno(4.7)$$
This follows from Eqs.~(3.11c) and~(4.3), expressed in terms of the
original classical amplitude~$x_0$.

The matrix products required are readily evaluated.  They are
$$\openup 2\jot
\eqalignno{N^{-1}U^{-1}VZ&={m_2x_0\over m_1+m_2}\,\left(\matrix{
\cos(\omega t)-\cos(\Omega t)\cr{}\cr
\cos(\omega t)+(m_1/m_2)\cos(\Omega t)\cr}\right)\ ,&(4.8)\cr}$$

$$\openup 2\jot
\eqalignno{N^TUN&={\omega\over\mu(m_1+m_2)[\cos^2(\Omega t)+\gamma^2
\sin^2(\Omega t)]}&\cr
&\qquad\times\left(\matrix{\mu^2+m_1^2[\cos^2(\Omega t)+
\gamma^2\sin^2(\Omega t)]&-\mu^2(1-\gamma^2)\sin^2(\Omega t)\cr{}&{}\cr
-\mu^2(1-\gamma^2)\sin^2(\Omega t)&\mu^2+m_2^2[\cos^2(\Omega t)
+\gamma^2\sin^2(\Omega t)]\cr}\right)\ ,&\cr&&(4.9)\cr}$$

$$\openup 2\jot
\eqalignno{N^T[U+(V-\Omega_2)U^{-1}V]Z&={\mu x_0\over m_1+m_2}\,\left(
\matrix{\omega\sin(\omega t)-\Omega\sin(\Omega t)\cr{}\cr
(m_2/m_1)\omega\sin(\omega t)+\Omega\sin(\Omega t)\cr}\right)\ ,&\cr
&&(4.10)\cr}$$

\noindent and
$$\openup 2\jot
\eqalignno{N^T[U+&(V-\Omega_2)U^{-1}(V-\Omega_2)]N=&\cr
&{\omega\over\mu(m_1+m_2)}\,\left(\matrix{{\displaystyle{m_1^2+\mu^2\left(
\cos^2(\Omega t)+{1\over\gamma^2}\sin^2(\Omega t)\right)}}&
{\displaystyle{-\left({1\over\gamma^2}-1\right)\mu^2\sin^2(\Omega t)}}\cr
{}&{}\cr
{\displaystyle{-\left({1\over\gamma^2}-1\right)\mu^2\sin^2(\Omega t)}}&
{\displaystyle{m_2^2+\mu^2\left(\cos^2(\Omega t)
+{1\over\gamma^2}\sin^2(\Omega t)\right)}}\cr}\right)\ .\cr
&&(4.11)\cr}$$
\vskip\baselineskip
\noindent{\bf C.~Expectation values and uncertainties.}\par\nobreak
The expectation values and uncertainties for position and momentum of
oscillator~\#1 in this state are obtained from the above matrices via
Eqs.~(3.22a), (3.22b), (3.23), and~(3.29).  They yield
$$\openup 2\jot
\eqalignno{\langle y_1\rangle&=x_0\,{m_2\over m_1+m_2}\,[\cos(\omega t)
-\cos(\Omega t)]\ ,&(4.12{\rm a})\cr
\sigma_y^{(1)}&=\sqrt{{\hbar\over2m_1\omega}\,{m_1\over m_1+m_2}\,\left(
1+{m_2\over m_1}[\cos^2(\Omega t)+\gamma^2\sin^2(\Omega t)]\right)}\ ,\qquad
&(4.12{\rm b})\cr
\langle p_1\rangle&=-m_1x_0\,{m_2\over m_1+m_2}\,[\omega\sin(\omega t)
-\Omega\sin(\Omega t)]\ ,&(4.12{\rm c})\cr
\hbox{and}\qquad\sigma_p^{(1)}&=\sqrt{{\hbar m_1\omega\over2}\,
{m_1\over m_1+m_2}\,\left[1+{m_2\over m_1}\left(\cos^2(\Omega t)+
{1\over\gamma^2}\sin^2(\Omega t)\right)\right]}\ .\qquad\qquad
&(4.12{\rm d})\cr}$$
The Gaussian probability distributions for position and momentum of
oscillator~\#1 follow immediately from these results.

The quantum behavior of oscillator~\#1 in this case resembles that in the
in the symmetric case, but exhibits a broader range of possibilities.  The
momentum and position expectation values satisfy
$\langle p_1\rangle=m_1\,d\langle y_1\rangle/dt$.  The position uncertainty
oscillates between the unperturbed value and a {\it smaller\/} value:
$$\sqrt{{\hbar\over2m_1\omega}\,{m_1+\gamma^2m_2\over m_1+m_2}}\le
\sigma_y^{(1)}\le\sqrt{\hbar\over2m_1\omega}\ .\eqno(4.13)$$
The momentum uncertainty compensates:
$$\sqrt{\hbar m_1\omega\over2}\le\sigma_p^{(1)}\le
\sqrt{{\hbar m_1\omega\over2}\,{m_1+\gamma^{-2}m_2\over m_1+m_2}}\ .
\eqno(4.14)$$
The uncertainty product oscillates between the Heisenberg minimum value and
a larger value---
$$\eqalignno{\sigma_y^{(1)}\sigma_p^{(1)}&={\hbar\over2}\,\sqrt{1+
{m_1m_2\over(m_1+m_2)^2}\left({1\over\gamma}-\gamma\right)^2\sin^2(\Omega t)\,
\left(1+{m_2\over m_1}\cos^2(\Omega t)\right)}\ .\cr&&(4.15{\rm a})\cr}$$
implying
$${\hbar\over2}\le\sigma_y^{(1)}\sigma_p^{(1)}\le{\hbar\over2}\,\sqrt{
1+{m_1m_2\over(m_1+m_2)^2}\,\left({1\over\gamma}-\gamma\right)^2}
\eqno(4.15{\rm b})$$
for $m_1\ge m_2$, or
$${\hbar\over2}\le\sigma_y^{(1)}\sigma_p^{(1)}\le{\hbar\over4}\,\left(
{1\over\gamma}+\gamma\right)\eqno(4.15{\rm c})$$
for $m_1\le m_2$---always satisfying the Heisenberg Uncertainty Principle.
However, unlike the symmetric case, for unequal-mass oscillators at
resonance the ``apparent loss of quantum nature,'' i.e., the quantum
squeezing of the uncertainties, can be extreme.  For a system with
$m_1\ll m_2$ and strong coupling, i.e., $\gamma\ll1$, the minimum value
of~$\sigma_y^{(1)}$ can be arbitrarily close to zero, the corresponding
maximum value of~$\sigma_p^{(1)}$ arbitrarily large.  Of course the
squeezing is oscillatory, with~$\sigma_y^{(1)}$ and~$\sigma_p^{(1)}$
returning to their unperturbed quantum values~$\sqrt{\hbar/(2m_1\omega)}$
and~$\sqrt{\hbar m_1\omega/2}$ in each cycle. 
\vskip\baselineskip
\centerline{\bf V.~~CONCLUSIONS}\par\nobreak
By virtue of its exact solubility, both classically and quantum-mechanically,
a pair of linearly coupled harmonic oscillators proves to be a useful toy model
for probing the interaction of a quantum system with its environment.  Even in
its ground state, this system displays effects of the coupling on the quantum
nature of an oscillator:  reduction of its position uncertainty below the
unperturbed quantum value; a compensating increase in its momentum uncertainty,
yielding an increase in the uncertainty product; breaking of the symmetry
between position and momentum variables, a consequence of the
position-dependent coupling.  These cannot be characterized as
finite-temperature effects.

The interaction of a quantum oscillator with a classical one can be
simulated via the coupled-oscillator pair in a quantum state with one
oscillator initially in its unperturbed ground state, the other initially
in a coherent or Glauber state incorporating classical behavior.  The
subsequent evolution of the wave function is calculated exactly, using
ordinary harmonic-oscillator propagators for the normal modes of the
system.  The reduced probability distribution for the position of the
initially quantum oscillator---a Gaussian distribution with time-dependent
expectation value and uncertainity---and its position and momentum
expectation values and uncertainties all follow from this wave function.
The expectation values can be characterized as a ``beat'' amplitude
between the normal modes.  The behavior of the uncertainties, i.e.,
the quantum character of the oscillator, can be quite complicated.
For oscillators with equal unperturbed frequencies, e.g., at resonance,
this behavior can be described as a time-dependent quantum squeezing:
The position uncertainty oscillates through values below the unperturbed
value; depending on the relative masses of the original oscillators and
the strength of the coupling, its lower bound can be arbitrarily close
to zero.  The momentum uncertainly oscillates through values larger than
its unperturbed value.  The uncertainty product oscillates through values
larger than the Heisenberg minimum value.  Naturally, the system never
actually violates the Uncertainty Principle.

Nonetheless, the behavior of the system is certainly reminiscent of the
longstanding suggestion that a quantum system coupled to a classical one
would ``radiate away'' its quantum nature.  In detail, though, it more
closely resembles the quantum squeezing encountered, e.g., in quantum
fields coupled to classical backgrounds~[8].  Although not explored in
the present calculations, more involved effects connecting the quantum
and classical worlds, such as decoherence, can be studied exactly and
analytically in the two-oscillator system.
\vskip\baselineskip
\centerline{\bf ACKNOWLEDGMENTS}\par\nobreak
The work of one of us (RMM) was supported by Saint Louis University Deans'
Tuition and Residence Life Scholarships, funded via the SLU Women's Council
Endowed Student Scholarship.
\vskip\baselineskip
\hrule
\item{[1]}C.~M{\o}ller, in {\it Les Theories Relativistes de la Gravitation,\/}
edited by A.~Lichnerowicz and M.~A.~Tonnelat (CNRS, Paris, 1962);
L.~Rosenfeld, Nucl.~Phys.~{\bf 40,\/} 353 (1963); B.~S.~DeWitt,
Phys.~Rep.~{\bf 19C,\/} 295 (1975); K.~Eppley and E.~Hannah,
Found.~Phys.~{\bf 7,\/} 51 (1977).

\item{[2]}D.~N.~Page and C.~D.~Geilker, Phys.~Rev.~Lett.~{\bf 47,\/} 979
(1981); B.~Hawkins, Phys.~Rev.~Lett.~{\bf 48,\/} 520 (1982);
L.~E.~Ballentine, Phys.~Rev.~Lett.~{\bf 48,\/} 522 (1982).

\item{[3]}E.~Schr\"{o}dinger, Naturwissenschaften~{\bf 14,\/} 664 (1926);
R.~J.~Glauber, Phys. Rev. Lett.~{\bf 10,\/} 84 (1963); Phys.~Rev.~{\bf 131,\/}
2766 (1963); J.~R.~Klauder and B.~S.~Skagerstam, {\it Coherent States,
Applications in Physics and Mathematical Physics\/} (World Scientific,
Singapore, 1985), pp.~3--24;  see also e.g., L.~D.~Landau and E.~M.~Lifshitz,
{\it Quantum Mechanics: Non-Relativistic Theory, Third Edition\/}
(Pergamon, Oxford, 1977), pp.~71--72.

\item{[4]}B.~L.~Schumaker, Phys.~Rep.~{\bf 135,\/} 317 (1986); W.-M.~Zhang,
D.~H.~Feng, and R.~Gilmore, Rev. Mod. Phys.~{\bf 62,\/} 867 (1990).

\item{[5]}D.~Stoler, Phys.~Rev.~D{\bf 1,\/} 3217 (1970); {\bf 4,\/} 1925
(1971); E.~Y.~C.~Lu, Lett.~Nuovo Cimento {\bf 2,\/} 1241 (1971); {\bf 3,\/}
585 (1971); J.~N.~Hollenhorst, Phys.~Rev.~D{\bf 19,\/} 1669 (1979);
J.~Grochmalicki and M.~Lewenstein, Phys.~Rep.~{\bf 208,\/} 189 (1991).

\item{[6]}A.~Messiah, {\it Quantum Mechanics, Volume~I\/} (Wiley, New York,
1958), pp.~448--451; R.~P.~Feynman, {\it Statistical Mechanics\/} (Benjamin,
Reading, Massachusetts, 1972), pp.~49--53.

\item{[7]}R.~P.~Feynman, Rev.~Mod.~Phys.~{\bf 20,\/} 367 (1948);
E.~W.~Montroll, Commun.~Pure Appl.~Math.~{\bf 5,\/} 415 (1952);
R.~P.~Feynman and A.~R.~Hibbs, {\it Quantum Mechanics and Path Integrals\/}
(McGraw-Hill, New York, 1965), pp.~62--63; L.~S.~Schulman, {\it Techniques
and Applications of Path Integration\/} (Wiley, New York, 1981), pp.~37--38.

\item{[8]}L.~P.~Grishchuk and Yu.~V.~Sidorov, Class.~Quantum Grav.~{\bf 6,\/}
L161 (1989); Phys.~Rev.~D{\bf 42,\/} 3413 (1990); L.~P.~Grishchuk, in {\it
Proceedings of the Sixth Marcel Grossmann Meeting,\/} edited by H.~Sato
and T.~Nakamura (World Scientific, Singapore, 1992); Class.~Quantum
Grav.~{\bf 10,\/} 2449 (1993);  Phys.~Rev.~D{\bf 50,\/} 7154 (1994);
{\bf 53,\/} 6784 (1996); Lect.~Notes Phys.~{\bf 562,\/} 167 (2001);
L.~Grishchuk, H.~A.~Haus, and K.~Bergman, Phys.~Rev.~D{\bf 46,\/} 1440 (1992);
M.~Gasperini and M.~Giovannini, Class.~Quantum Grav.~{\bf 10,\/} L133 (1993);
Phys.~Lett.~{\bf 301B,\/} 334 (1993); S.~Bose and L.~P.~Grishchuk,
Phys.~Rev.~D{\bf 66,\/} 043529 (2002).
\vfill\eject\end